\documentstyle[aps,epsfig,multicol,fancyheadings]{revtex}
\begin{document}

\title{Scaling behaviour in the fracture of fibrous materials}

\author{I. L. Menezes-Sobrinho $^1$, J. G. Moreira$^1$ and 
A. T. Bernardes$^2$} 

\address{$^1$ Departamento de F\'{\i}sica, Instituto de Ci\^encias 
Exatas, Universidade Federal de Minas Gerais\\
C.P. 702, 30123-970, Belo Horizonte, MG, Brazil\\
$^2$ Departamento de F\'{\i}sica, Universidade Federal 
de Ouro Preto\\
Campus do Morro do Cruzeiro, 35400-000 Ouro Preto, MG, Brazil}

\date{\today}
\maketitle

\begin{abstract}
We study the existence of distinct failure regimes in a model for fracture 
in fibrous materials. We simulate a bundle of parallel fibers under uniaxial
static load and observe two different failure regimes: a catastrophic and 
a slowly shredding. In the catastrophic regime the initial deformation produces 
a crack which percolates through the bundle. In the slowly shredding regime 
the initial deformations will produce small cracks which gradually weaken the 
bundle. The boundary between the catastrophic and the shredding regimes is 
studied by means of percolation theory and of finite-size scaling theory. 
In this boundary, the percolation density $\rho$ scales with the 
system size $L$, which implies the existence
of a second-order phase transition with the same critical exponents as 
those of usual percolation.
\end{abstract}

\vskip 2truecm
\noindent{Pacs: 62.20.Mk; 05.40.+j; 64.60.Ak}
\begin{multicols}{2}[]
\narrowtext

\section{Introduction}
	
The nature of fracture of non-homogeneous materials is an important 
problem in material science research.
Computer simulation of the fracture phenomenon are very useful since
the analytical approachs is very difficult to perform.
This difficulty arises from the non-uniform character of the material
and their discrete nature, which are fundamental ingredients for understanding
the rupture process \cite{hans-roux}. 
Usually, computer simulation in these materials gives interesting results, 
however the high degree of correlations between the constituents  
leads to a high computational cost. Bundles of unidirectional
fibers form a system with low degree of correlations allowing the 
fracture process be simulated in a large scale. 

The study of fibrous materials is not recent, as one can find in the work
of Daniels \cite{daniels}, who in 1941 studied the rupture of a bundle of 
fibers with a known probability distribution of strength. Recently, Hansen 
and Hemmer \cite{hansen-hemmer} studied the distribution $H(S)$ of the sizes 
$S$ of burst avalanches, i. e., an instantaneous propagation of a crack.
They found a power-law behaviour: $H(S)\sim S^{-\alpha}$ with the
exponent $\alpha$ depending on how the load is shared between
the fibers. For global load sharing, where the load is shared equally
among non breaking fibers, they obtained $\alpha = 2.5$. 
For local load sharing, when a fiber breaks its load is shared among 
nearest-neighbours no breaking fibers, they obtained  $\alpha = 4.5$. 
Those results have been obtained for a
one-dimensional lattice of fibers, that is, the load sharing is the
only correlation between the fibers. Other approaches have been 
introduced to discuss this problem \cite{duxbury,zhang}.
The role of the homogeneous support matrix on the failure of composite 
materials has also been discussed by some authors
\cite{phoenix,zhou,Fat96}.

A model of a bundle of unidirectional fibers, which takes into account
external parameters like temperature and velocity of traction, has been 
proposed in 1994 by Bernardes and Moreira \cite{bm94}. In this model, the 
correlations between the fibers are present through the probability of rupture
of a fiber, which depends on the number of unbroken nearest neighbouring fibers. 
A cascade mechanism - inspired on models for avalanches - is used to propagate 
cracks through  the material: when a fiber breaks, its neighbours are visited
and can break too. In this model, all unbroken fibers 
have the same deformation, i.e, one has global load
sharing. However, the cascade mechanism 
introduces a local effect. In a subsequent work\cite{bm95}, 
the dependence of the frequency of cracks with the crack sizes were used  
to determine the failure regimes. Two basic regimes
were discussed: a regime where cracks of the size of the system were present 
and another one where only small cracks appeared. Those regimes were identified, 
respectively, with the brittle and ductile failure regimes. The criterion used 
to distinguish one regime from another was based on self-organized criticality, 
i.e, in the brittle-ductile transition region, cracks of all sizes were present.
However, they did not take into account finite size effects in their
analysis, which are very important in this type of process. In fact,
a finite-size scaling analysis should be performed, in order to
guarantee a better definition of the failure regimes.

The aim of the present paper is to introduce a criterion which defines the 
failure regimes on fibrous materials. When a fiber bundle breaks, two regimes 
can be observed: a catastrophic regime, when a sufficiently large number of
fibers are simultaneously broken, and a slow process of successive rupture of 
fibers, here called shredding regime. The first regime occurs at low 
temperatures and/or high strains and are similar to a brittle fracture. 
It is characterized by the fact that an initial deformation
produces a large crack which percolates through the bundle.
The shredding regime occurs for higher temperatures and/or lower strains,
and is similar to the ductile regime. In this case, the first deformations
produce small cracks which weaken the bundle and thus cause its failure.
The criterion is implemented by considering the static failure of a 
modified version of the model introduced by Bernardes and Moreira \cite{bm94}. 
A second order phase boundary between two regimes 
is found for a given strain. A finite-size scaling analysis is used 
to determine the critical temperature and exponents.

\section{The Model}
	
The model for the fibrous material here discussed
consists of a bundle of $N_0=L\times L$ parallel 
fibers with a cross-section forming a triangular lattice. 
Each fiber has the same 
elastic constant $k$, and they are fixed at both ends to parallel 
plates. One plate is fixed and the other plate can be pulled
by an external force. 
When the bundle is pulled by a force $F$, all fibers undergo the 
same linear deformation $z = F/Nk$, where $N$ is the number of unbroken fibers.
We assume that a fiber has a failure probability which increases with the 
deformation $z$.
When this deformation reaches a critical value $z_c$,
the breaking probability of an isolated fiber is equal to one. 
When the bundle has a deformation $z$, a fiber $i$ has a failure probability 
related to its elastic 
energy and to the number of unbroken neighbouring fibers $n_i$, given by 

\begin{equation}
\label{fp0}
P_i(\delta)=
{z/z_c\over n_i+1}\exp \left({(kz^2/2)-(kz_c^2/2)\over K_BT}\right) ~~,
\end{equation}

\noindent Defining the strain of the material as $\delta=z/z_c$ and  
the normalized temperature as $t=K_BT/E_c$, where $T$ is the absolute 
temperature, $E_c=kz_c^2/2$ is the critical elastic energy and $K_B$ 
is the Boltzman constant, we can rewrite the failure probability as 

\begin{equation}
\label{fp}
P_i(\delta)=
{\delta\over n_i+1}\exp \left({\delta ^2 - 1\over t}\right)~~,
\end{equation}

This definition of the failure probability
is different from that used by Bernardes and Moreira\cite{bm94},
since now we have introduced $\delta$ as a multiplicative factor to 
impose that, for $\delta = 0$, $P_i(\delta)=0$.

The static failure of a fiber bundle is produced by applying a constant force 
$F_0$ to the bundle, for example, by hanging a weight on the moving plate. 
The initial strain  of the bundle is given by

\begin{equation}
\delta_0 = { z_0 \over z_c} = { F_0 \over {N_0 k z_c} }
\end{equation}

\noindent The simulation of the rupture process proceeds as follows.
At each time step of the simulation, we randomly choose a set of 
$N_q (= q N_0)$ unbroken fibers, where the number $q$ represents a percentage 
of fibers and it allows us to work with any system size. 
So, differently of an Ising model, where all the sites are ``tested''
at each time step, in our model only a number $N_q$ of randomly chosen unbroken
fibers  are tested. It represents the continuous growth of
the bundle due to the continuous traction.
For each chosen fiber, we evaluate the probability of rupture, using Eq. 1, 
 and compare it with a random number in the interval 
[0,1). If the random number is less than the failure probability, the fiber 
breaks. To simulate the load spreading,
the same process is repeated for all neighbouring unbroken fibers.
The failure probability of these neighbouring fibers 
increases due to the decreasing of $n_i$ and a cascade of breaking 
fibers may begin. This procedure describes the propagation of a crack 
through the fiber bundle, which occurs in all directions 
perpendicular to the force applied to the system. 
The cascade process stops when the test of the probability does not 
allow the rupture of any other fiber on the border of the crack or
when the crack meets another already formed crack. This collision leads to
the fusion of cracks, and it is the mechanism to explain the rupture of the 
material in the shredding regime. 
The same cascade propagation is attempted by choosing another fiber of the 
set  $N_q$. 
After all the $N_q$ fibers have been tested, the strain is increased if some 
fibers have been broken. 
This new strain is the same for all the remaining unbroken fibers.
Since the force is fixed (the weight hung on the bundle), the greater the 
number of broken fibers, the larger is the strain on the fibers, and the higher 
is the failure probability.
Then, other set of $N_q$ unbroken fibers are chosen and the rupture process 
restarts. This process stops when all the fibers are broken, i.e, 
the bundle breaks apart. In this model, a combination of local and global 
load sharing occurs. That is, after a fiber breaks, a cascade may begin which 
simulates 
the local load sharing. When the cascade process stops, the stress is 
distributed equally between all unbroken fibers which is the global load 
sharing.

\section{RESULTS}

The failure probability (Eq. \ref{fp}) can be written as

\begin{equation}
\label{fp1}
P_i(z)= {\Gamma (t,\delta) \over {(n_i +1)}  }~~~~~,
\end{equation}

\noindent where we introduce the parameter $\Gamma (t,\delta)$ defined as

\begin{equation}
\label{po}
\Gamma (t,\delta) = \delta\exp \left({\delta ^2 - 1} \over t \right)~~.
\end{equation}

For a triangular lattice (with coordination number 6) and 
$\Gamma (t,\delta) \geq 6$, the rupture of any fiber induces
the rupture of the whole bundle, i.e., the bundle 
 breaks with just one crack. Obviously, this crack forms a cluster
 which percolates through 
the entire system.

We can define the density of the percolation crack as

\begin{equation}
\label{rho}
\rho = {N_{pc}\over N_0}~~,
\end{equation}

\noindent where $N_{pc}$ is the number of broken fibers belonging
to the percolating crack. Thus, when $\Gamma (t,\delta) \geq 6$, we
have $\rho = 1$.
On the other hand, as it has been observed in previous
works \cite{bm95,ijmpc98}, for higher temperatures and/or lower
strains, the fracture of the bundle is caused by many small
cracks, none of then large enough to percolate through the system.
Thus, for a fixed temperature, if one starts with
a large enough strain $\delta_0$ and one decreases it, the system goes from 
a regime 
where $\rho = 1$ to another regime where $\rho \rightarrow 0$.
This behaviour is the same as the one encountered in the percolation problem. 

Figure 1 shows the density of the percolation cluster $\rho$ versus the 
initial strain $\delta_0$, for two different 
temperatures. As one sees, $\rho=1$ for high values of $\delta_0$,
and jumps to zero for low enough value of $\delta_0$.
So, we may assume that, for a fixed temperature,
there is a critical value $\delta_{0c}$ above which one observes
a percolation crack, and below which there is no
percolation at all. Another interesting feature that one can observe in 
Figure 1 is that, if one substitutes into Eq. \ref{po} the values of $t$ 
and $\delta_{0c}$
corresponding to the transition region 
($\delta _0 \sim 1.18$ for $t=1.0$ and $\delta _0 \sim 1.37$ for $t=4.0$),
we get for both instances $\Gamma (t,\delta) \sim 1.73$. The fundamental 
reason for
obtaining this value wil be explained below.
 
In contrast to that described above, the same behaviour does not
occur when we keep $\delta_0$ fixed  and change the temperature.
Figure 2 shows the results obtained for 
the density of the percolation cluster $\rho$ versus temperature
$t$, for $\delta_0=1.4$. We observe that, initially, $\rho$
decreases as the temperature $t$ increases, and around $t \sim 4.5$,
the value of $\rho$ seems to go to zero. However, an additional increase
in the temperature will revert the process and a minimum appears.
Note that at the point of minimum again 
($\Gamma(t=4.5, \delta=1.4) \sim 1.73)$, which is the same value reported above.
For low temperatures ($t < 2.0$), 
when a fiber breaks, the probability is so high 
that this rupture initiates a cascade which breaks the
whole bundle. By increasing the temperature, a number of
small cracks are formed, inhibiting the
formation of a percolating cluster and the density $\rho$ decreases. 
However, all those processes occur in the first step of the simulation
when $N_q$ attempts to break the bundle are performed. 
Thus, for $t < 4.5$, the bundle has been broken due to the crack which
percolates the system during the first $N_q$ attempts to break it. 
For $t > 4.5$, all
the first $N_q$ attempts do not succeed to generate a crack which
percolates the bundle. However,
some fibers have been broken and cracks were formed.
In the second step of the simulation,
a new value of $\delta$ is used (higher than $\delta_0$)
and a new set of $N_q$ trials are chosen. But now one has a higher
value for $\Gamma (t,\delta)$  therefore it 
is easier to produce a large crack which percolates the bundle. 
By increasing the temperature, a smaller number of fibers are 
broken in the first $N_q$ attempts, and then, in the second step, the 
are more unbroken fibers and therefore the density $\rho$ increases, thus 
forming a minimum in the graph of Figure 2.

In fact, we can assume that 
there is a critical value for $\Gamma (t,\delta)\sim 1.73$ that
defines the transition between two regimes. In the first one,
a catastrophic fracture occurs due to the first attempt to break the bundle,
while in the second case the rupture of the bundle occurs due
to the formation of small cracks, which weaken the bundle.
A percolating crack may also occur in the second case, however
the  fracture dynamics is given by the weakening of the bundle 
not by the catastrophic propagation of a crack. 

In order to consider the present model in the context of percolation
theory, we shall use the parameter $\Gamma$ as an arbitrary parameter
without regarding it as a function of the strain $\delta$ and temperature 
$t$. Within the percolation point of view, we map the original model into
a triangular lattice where the empty sites corresponds to the unbroken
fibers. The parameter $\Gamma$ is the analog to the percolation probability.
The algorithm for the mechanism of fracture is mapped into the following
algorithm for the percolation problem.
An empty site (unbroken fiber) $i$ is chosen at random;
Its occupation (failure) probability $P_i$ is calculated by dividing $\Gamma$ 
by the number of its neighbouring empty sites (unbroken fibers) plus one.
This probability $P_i$ is compared with a random number $r\in [0,1)$;
If $P_i > r$, the site is occupied (the fiber is broken) and a cluster
(crack) can be formed, i.e, an empty neighbouring site (an unbroken 
neighbouring fiber) is randomly chosen and the process are repeated;
Otherwise, another site, on $N_q$ in total, is chosen.

When a cluster is formed, we test if it percolated through the system.
If it does, we calculate the density $\rho$ of the percolating cluster. 
Figure 3a shows the results obtained for several system sizes. 
Two regions are separated by the transition point $\Gamma_c$.
The larger the system size, more 
clearly is the transition between those two regions. Observe
in the detail, shown in Figure 3b, that a 
second order phase transition takes place at $\Gamma_c= 1.733(1)$.
This implies that, at that point, clusters of all sizes  should be present,
as confirmed by the results shown in Figure 4. In this figure, the results
 have been obtained for a system
size $L=5000$ ($2.5\times 10^7$ fibers)
and averaged over 1000 samples (it took nearly 24h
on a Sun Enterprise 8GB computer) which gives the following power law  

\begin{equation}
\label{pl}
H(S)\sim S^{-\tau}~~~,
\end{equation}

\noindent  where $\tau=2.037\pm 0.007$. A finite size analysis can be 
performed by plotting 
$\tau(L)$ as a function of $L^{-1/\nu}$, where $\nu$ is the exponent related
to the divergence of the correlation length at the transition. We 
tested several value of $\nu$ and the best linear fitting were obtained 
for $\nu =4/3$, as shown in Figure 5. 
This value corresponds to the exact exponent $\nu$
for percolation at $d=2$. 
The value of the exponent $\tau$ for an infinite lattice is, then, evaluated
to be $\tau(\infty) = 2.05 \pm 0.01$, in an excellent agreement with the
theoretical value, $\tau _{\infty} = 2.055$ \cite{stauffer}.

In order to check if our problem belongs to the same universality as the
percolation problem, we have done a finite-size 
scaling analysis by assuming the scaling law \cite{stauffer}

\begin{equation}
\label{ref3}
\rho(\Gamma,L)= L^{-\beta / \nu}\psi\left(\epsilon L^{1/ \nu} \right)~~,
\end{equation}
	
\noindent where

\begin{equation}
\epsilon= {\left| 1 - {\Gamma \over \Gamma_c }\right| }~~,
\end{equation}  

\noindent $\psi~$ is a universal function of $\epsilon L^{1/\nu}~$ only,
and $\beta~$ and $\nu~$ are the critical exponents for the infinite lattice.
Figure 6 shows the finize-size scaling plot 
$\rho L^{\beta/\nu}$ versus $\epsilon L^{1/ \nu}$
for nine sizes of $L$.
We have used $\Gamma_c = 1.733$, $\nu=4/3$ and the best value of $\beta$
which validates Eq. \ref{ref3} is $\beta=0.14$. This value is also in
an excellent agreement with the known value for the usual percolation.

Now, returning to the original fracture model, we use Eq. \ref{po}
to obtain the critical temperature $t_c$ in terms of the critical parameter 
$\Gamma_c$ and of the initial strain $\delta_0$

\begin{equation}
t_c={\delta_0^2-1\over{\ln(\Gamma_c)-\ln(\delta_0)}}.
\end{equation}

Using this expression we plot the fracture regimes diagram, in
the temperature $t$ versus 
the initial strain $\delta_0$ plane, depicted in Figure 7. 
Two fracture regimes are separated by a second order transition line. 
In region {\bf C} the fracture is catastrophic and in region {\bf S} we have 
the  shredding 
regime. Note that the catastrophic regime only occurs for $\delta_0>1$ and 
for low temperatures. In this figure, the solid line corresponds to 
the analytical results, and the points were obtained by simulations.

\section{Conclusions}

In conclusion, we have studied a model for fracture in fibrous materials
in (2+1)-dimensions and shown the existence of two failure regimes:
the catastrophic regime, where the initial deformation produces a single
crack which percolates through the bundle; and the 
slowly shredding regime, where the initial deformation produces small 
cracks which gradually weaken the bundle.
By using percolation theory and finize-size scaling arguments, we were able 
of finding the transition line between these regimes. Our results indicate that 
this transition is of second order. Finally, we have shown that this model 
belongs to the same universality class as the percolation problem. 
 
\noindent {\bf Acknowledgements} We thank H. J. Herrmann and J. A. 
Plascak for fruitful discussions and suggestions. 
C. Moukarzel and M. Continentino did important
suggestions about the approach through percolation theory.
We also thank O. F. de Alcantara Bonfim for helpful criticism of the manuscript. 
One of us (ATB) acknowledges 
the kind hospitality of the Depto de F\'{\i}sica, UFMG. We also acknowledge 
CNPq and FAPEMIG (Brazilian agencies) for financial support. Most of 
our simulations have been performed on the Sun Enterprise 8GB computer of 
the CENAPAD MG/CO.

\newpage

\begin{figure}[f]
\centerline{\psfig{file=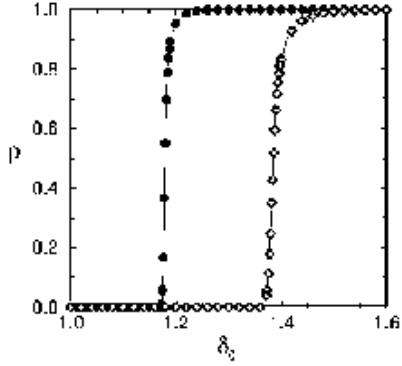,width=11cm,angle=0}}
\caption{Density of the percolating cluster $\rho$ versus the initial strain $\delta_0$
for two different temperatures: $t=1.0$ (filled circles) and 
$t=4.0$ (open diamonds). The system size is
$L= 1000$ and the data were averaged over 1000 statistically independent
samples.} 
\label{fig1} 
\end{figure}

\begin{figure}[f]
\centerline{\epsfig{file=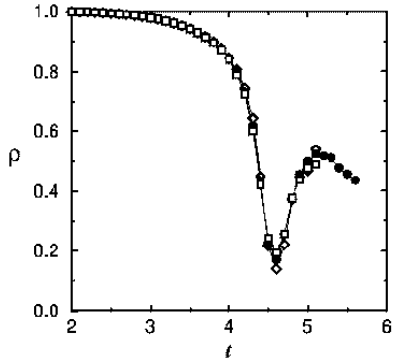,width=11cm,angle=-0}}
\caption{Density of the percolating cluster $\rho$ versus temperature for an initial 
strain $\delta_0=1.4$ and three different system sizes: $L=800$ (open
squares); $L=900$ (filled circles) and $L=1000$ (open diamonds). 
The data were averaged over 1000 statistically independent
samples.} 
\label{fig2} 
\end{figure}

\begin{figure}[f]
\centerline{\epsfig{file=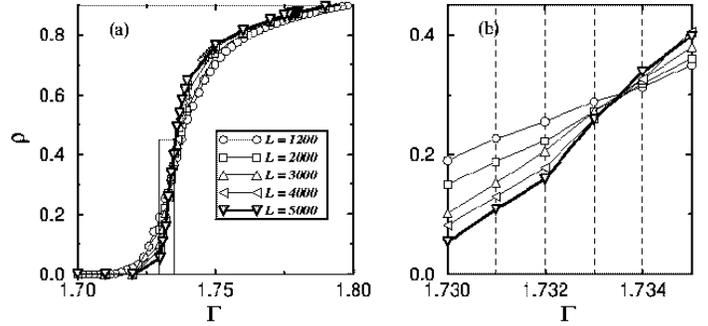,width=11cm,angle=-0}}
\caption{(a) Density of the percolating cluster $\rho$ versus $\Gamma$ for five 
different 
system sizes; (b) Zoom of the region corresponding to the small box
drawn in the left plot, showing more clearly the crossing of the curves.
The data were averaged over 1000 statistically independent
samples.} 
\label{fig3} 
\end{figure}

\begin{figure}[f]
\centerline{\epsfig{file=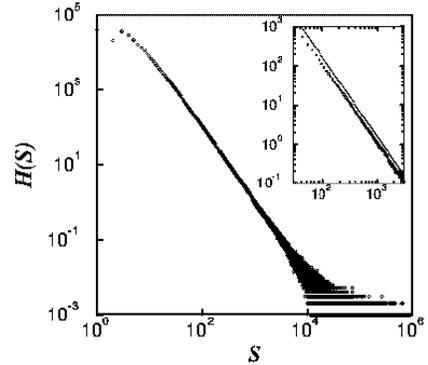,width=11cm,angle=-0}}
\caption{Log-log plot of the averaged number of cracks $H(S)$ versus
the crack size $S$ for $L=5000$ at $\Gamma_c=1.733$. The points have been 
obtained by averaging over 1000 statistically independent samples. 
The data show a  power law behaviour (expected at
the criticality) with exponent $\tau=2.037 \pm 0.007$.
The insert shows a detail of the whole set. The solid line
in this insert has exponent 2.037.} 
\label{fig4}
\end{figure}

\begin{figure}[f]
\centerline{\epsfig{file=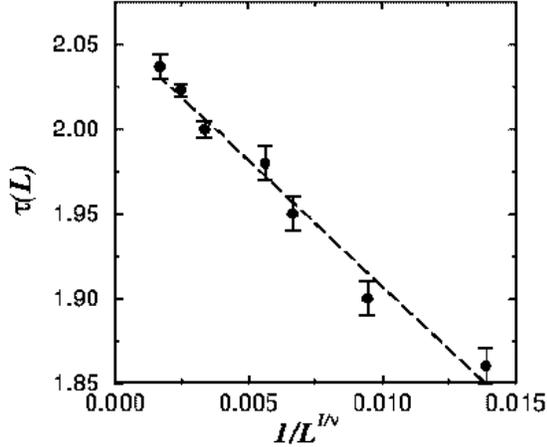,width=14cm,angle=-0}}
\caption{Estimate of the value of $\tau_{\infty}$.
We plot the value of $\tau(L)$ versus $L^{-1/\nu}$ with $\nu=4/3$. 
A linear regression
has been performed, giving $\tau_{\infty}=2.05 \pm 0.01$.}
 
\label{fig5} 
\end{figure}

\begin{figure}[f]
\centerline{\epsfig{file=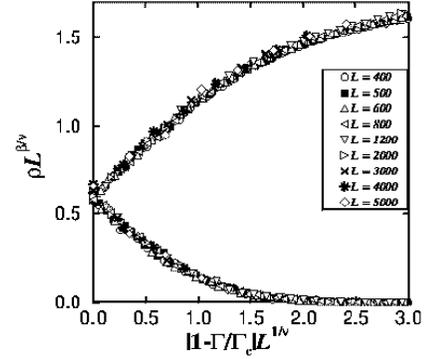,width=11cm,angle=-0}}
\caption{Plot of the scaling relation 
$\rho L^{\beta/\nu}$ versus $\epsilon L^{1/\nu}$ for nine 
system sizes (provided in the legend) with $\Gamma_c=1.733, ~\beta=0.14$ 
and $\nu=4/3$.} 
\label{fig6} 
\end{figure}

\begin{figure}[f]
\centerline{\epsfig{file=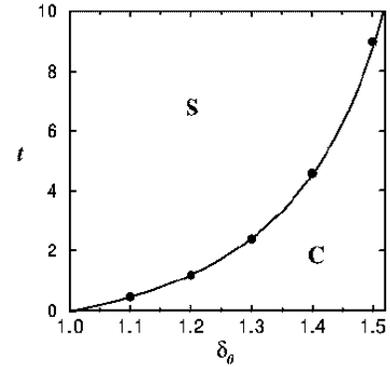,width=11cm,angle=-0}}
\caption{Fracture regimes diagram of the temperature 
$t$ in function of 
the initial strain $\delta_0$, where {\bf C} represents the catastrophic 
regime and {\bf S} represents the shredding regime. Solid line represents the
theoretical curve and filled circles represent the data obtained
in our simulations.} 
\label{fig7}
\end{figure}
\end{multicols}
\widetext

\end{document}